\documentclass{article}
\usepackage[super,compress]{cite}
\usepackage{graphicx}
\usepackage{amsmath,amssymb}
\begin{document}

\markboth{L. \'A. Somlai, M. Vas\'uth }
{The effect of the cosmological constant}

%\catchline{}{}{}{}{}
%
%%%%%%%%%%%%%%%%%%%%%%%%%%%%%%%%%%%%%%%%%%%%%%%%%%%%%%%%%%%%%%%%%%%%

\title{The effect of the cosmological constant on a quadrupole signal in the
linearized approximation}

%\author{L. \'A. Somlai  \footnote{Institute for Particle and Nuclear Physics, Wigner Research Centre for Physics, Hungarian Academy of Sciences,
%	Konkoly-Thege Mikl\'{o}s \'{u}t 29-33., H-1121 Budapest, Hungary}}
%\address{somlai.laszlo@wigner.mta.hu}
\author{\small
{L\'{a}szl\'{o} \'{A}bel Somlai}\thanks{email: somlai.laszlo@wigner.mta.hu}  and M\'{a}ty\'{a}s Vas\'{u}th\thanks{email: vasuth.matyas@wigner.mta.hu}
\\ %EndAName
\small WIGNER RCP, RMKI \\
\small  H-1121 Budapest, Konkoly Thege Mikl\'os \'ut 29-33.\\
\small Hungary
\date{}
}
\maketitle

\begin{abstract}
In this study the effects of a non-zero cosmological constant $\Lambda$ on a quadrupole gravitational wave signal are analysed. The linearized approximation of general relativity was used, so the perturbed metric can be written as the sum of $h^{GW}$ gravitational waves and $h^{\Lambda}$ background term, originated from $\Lambda$. The $\Lambda h^{GW}$ term was also included in this study. To derive physically relevant consequences of $\Lambda\neq0$ comoving coordinates are used. In these coordinates, the equations of motion are not self-consistent so the result of the linearized theory have to be transformed to the FRW frame. The luminosity distance and the same order of the magnitude of frequency in accordance with the detected gravitational waves were used to demonstrate the effects of the cosmological constant.
\end{abstract}

\section{Introduction}	

In the last decades, the cosmological and astrophysical observations have testified that the Universe is accelerating at late eras, see {\it e.g.} \cite{key-15}. This effect is indicated by {\it i.a.} the detailed study of the cosmological microwave background radiation\cite{key-12,key-13} and the gas fraction in X-ray luminous galaxy clusters \cite{key-14}. Among the potential models which describe the acceleration, most of the observations prefer the presence of Cold Dark Matter with positive cosmological constant. The $\Lambda CDM$ concordance model become our current cosmological picture. According to the present observations, the two dark components of the model account for ~95\% of the energy content of the Universe \cite{key-16}.

The year 2016 has been very intense in gravitational physics, the LIGO-Virgo collaboration has announced the direct detection of gravitational waves (GWs) for the first time \cite{key-6,key-9} opening an emerging research field of GW astronomy. The observed waveforms also fit to the current $\Lambda CDM$ cosmological model and to Einstein's prediction of GWs \cite{key-8}. In consequence, the physical properties of the detected GWs can be used in the study of perturbative effects, just like the presence of  a non-zero cosmological constant $\Lambda$ and its theoretical consequences \cite{key-9,key-16} as well.

To study these small effects, especially the effect of $\Lambda$ one has two major  methods. The first one is the gauge-invariant wave extraction method with the perturbative treatment of the metric tensor. In this approximation the $g_{\mu\nu}$ metric tensor can be written as the sum of the flat metric and a perturbatively consistent $h_{\mu\nu}$ term which contains the cosmological constant. The other is the investigation of the component of the Weyl tensor, the so-called tetrad formalism. Both methods provided relevant improvements in this field \cite{key-2,key-5}.

In the case of the metric perturbation treatment the leading effect in order to reach a physically significant coordinate system is the Friedmann--Robertson--Walker coordinate transformation which involves $\mathcal{O}(\sqrt{\Lambda})$ terms. For a consistent linearization of the Einstein equations the so-called $\Lambda$ gauge is used which results other coordinate transformation involving $\mathcal{O}(\Lambda)$ terms. The resulting equations of motion can be rewritten order by order of $\Lambda$ which also generates $\mathcal{O}(\Lambda)$ terms. Due to the modified gauge condition, there is no further symmetry to reduce the degrees of freedom, so massive excitation will appear at the same $\mathcal{O}(\Lambda)$ order.

This paper is organized as follows. In Sec. \ref{sec:2} the key ideas of the linearization method are discussed, including the gauge fixing process. In Sec. 3 the effects of a non-zero $\Lambda$ on the quadrupole signal are derived and, in particular, its dependence on different distance scales are analyzed. Sec. 4 contains the summary of our results.

\section{Linearization and gauge fixing}
\label{sec:2}

In this section a brief summary of the key ideas will be considered. A more detailed discussion can be found in \cite{key-1,key-2} and the references therein.

\subsection{Linearization}

Einstein field equations have the following form in the presence of the $\Lambda$ cosmological constant

\begin{equation}
R_{\mu\nu}-\frac{1}{2}g_{\mu\nu}R+\Lambda g_{\mu\nu}=-\kappa T_{\mu\nu}\ ,
\label{eq:einstein egyenlet, nem perturbalt}
\end{equation}
where $\Lambda>0$, $R_{\mu\nu}$ is the Ricci tensor for $g_{\mu\nu}$, $R$ is the Ricci scalar, $\kappa T_{\mu\nu}$ is the source term and $T_{\mu\nu}$ is the stress-energy tensor. In this work we are considering $\kappa T_{\mu\nu}=0$ until subsection \ref{subsec:pert}. In the following, the linearized theory is considered where the metric tensor can be written as

\begin{equation}
g_{\mu\nu}=\eta_{\mu\nu}+h_{\mu\nu}=\eta_{\mu\nu}+h_{\mu\nu}^{\Lambda}+h_{\mu\nu}^{GW}
\label{eq:metrika sorfejtes}
\end{equation}

$\eta_{\mu\nu}$ is the Minkowski metric, $h_{\mu\nu},h_{\mu\nu}^{\Lambda},h_{\mu\nu}^{GW}\ll1$,
$h_{\mu\nu}^{GW}$ can be interpreted as gravitational waves and $h_{\mu\nu}^{\Lambda}$ is the background perturbation. This theory is gauge invariant under the coordinate transformation which will be explained in the next section. In the next subsections two different gauge choices will be discussed.

\subsection{Lorenz gauge}

The usual option is the use of the Lorentz gauge where the gauge fixing conditions
have the following form \cite{key-4}

\begin{equation}
\partial_{\mu}h_{\,\nu}^{\mu}=\frac{1}{2}\partial_{\mu}h\rightarrow\partial_{\mu}\tilde{h}_{\,\nu}^{\mu}=0
\end{equation}
where $h=h_{\,\sigma}^{\sigma}$ and $\tilde{h}_{\mu\nu}=h_{\mu\nu}-\frac{1}{2}\eta_{\mu\nu}h$
is the trace reverse of $h_{\mu\nu}$. In this gauge the equations of motion (EoMs) are written as

\begin{equation}\label{LorentzEoM}
\Box\left(h_{\mu\nu}-\frac{1}{2}\eta_{\mu\nu}h\right)+2\Lambda h_{\mu\nu}=-2\Lambda\eta_{\mu\nu}.
\end{equation}

The EoMs have a residual gauge symmetry if, and only if the $\Lambda h_{\mu\nu}$ in the l. h. s. of (\ref{LorentzEoM}) is omitted. In this case one can use the linear coordinate transformations

\begin{equation}
x^{\mu}\rightarrow x'^{\mu}=x^{\mu}+\xi^{\mu}\,\,\,if\,\,\,\Box\xi^{\mu}=0
\label{eq:residualSymm in Lorenz}
\end{equation}
to set the transverse traceless gauge.

\subsection{$\Lambda$-gauge}

In order to study the effect of $\Lambda$ on the linearized theory
a different gauge choice is considered \cite{key-3}. One can introduce the so-called
$\Lambda$-gauge, which is defined by the gauge conditions

\begin{equation}
\partial_{\mu}\tilde{h}_{\,\nu}^{\mu}=-\Lambda\eta_{\mu\nu}x^{\mu}\ .
\label{eq:Lambda k=0000E9nyszer}
\end{equation}
The EoMs have the following form
\begin{equation}
\Box\left(h_{\mu\nu}-\frac{1}{2}\eta_{\mu\nu}h\right)+2\Lambda h_{\mu\nu}=0.\label{eq:Lambda mozgegyenlet}
\end{equation}

The $\Box\xi^{\mu}=-\Lambda\xi^{\mu}$ residual coordinate transformations are not a symmetry of the EoMs, they only have the global Lorenz transformation symmetry. Therefore, these transformations cannot be used to remove degrees of freedom. If the $\Lambda h_{\mu\nu}$ terms are omitted, the following coordinate transformations connect the two different gauge choices

\begin{equation}
x^{\mu}\rightarrow x'^{\mu}=x^{\mu}+\xi^{\mu}=\left(1-\frac{\Lambda}{12}x^{2}\right)x^{\mu}.
\label{eq:atvaltas x->x' kozott}
\end{equation}

This means that we have to select such a coordinate system in the Lorenz gauge which suitably fulfill the EoMs after the transformation (\ref{eq:atvaltas x->x' kozott}), that is in the $\Lambda$-gauge. Other coordinate systems, which are connected with the chosen system by the residual symmetry (\ref{eq:residualSymm in Lorenz}), will produce off-shell results in the $\Lambda$-gauge. If the $\Lambda h$ terms are retained, there is no residual symmetry and the degrees of freedom cannot be reduced, so massive excitation will appear. In this paper we will not study the effects of these massive excitations. For a more detailed discussion see \cite{key-10}.

\subsection{Perturbation}
\label{subsec:pert}

Recall that metric perturbations can be written as $h_{\mu\nu}=h_{\mu\nu}^{\Lambda}+h_{\mu\nu}^{GW}$. To study the $\Lambda$-order solution, first, we can solve Eqs. (\ref{eq:Lambda k=0000E9nyszer}) and (\ref{eq:Lambda mozgegyenlet}) for $\tilde{h}^{\Lambda}$

\begin{equation}
\tilde{h}_{\mu\nu}^{\Lambda}=-\frac{\Lambda}{18}\left(4x_{\mu}x_{\nu}-\eta_{\mu\nu}x^{2}\right)\ .
\label{eq:h megold=0000E1sa Lambda gaugeban}
\end{equation}
The solution is unique if we require that $h^{\Lambda}$ is proportional to $\Lambda$ and involves the coordinates $x^{\mu}$ only. Moreover, we can expand $\tilde{h}^{GW}$ as

\begin{equation}
\tilde{h}_{\mu\nu}^{GW}=\tilde{h}_{\mu\nu}^{(0)\,GW}+\tilde{h}_{\mu\nu}^{(1)\,GW}+\mathcal{O}(\Lambda^{2})
\label{eq:h kifejtve Lamb =0000E9s GW-vel}
\end{equation}
where a superscript refers to the order in $\Lambda$. The EoMs can be written in terms of $\tilde{h}_{\mu\nu}^{GW}$ and $\tilde{h}_{\mu\nu}^{\Lambda}$ separately. For $\tilde{h}^{\Lambda}$ the relavant equations are

\begin{eqnarray}
\partial_{\mu}\tilde{h}_{\,\,\nu}^{\Lambda\mu} & = & -\Lambda\eta_{\mu\nu}x^{\mu}\\
\Box\tilde{h}_{\mu\nu}^{\Lambda} & = & 0\nonumber
\end{eqnarray}

In the presence of source terms the EoMs for $h_{\mu\nu}^{GW}$ are

\begin{equation}
\Box h_{\mu\nu}^{(0)\,GW}+\Box h_{\mu\nu}^{(1)\,GW}+2\Lambda h_{\mu\nu}^{(0)\,GW}+\mathcal{O}\left(\Lambda^{2}\right)=-\kappa T_{\mu\nu}\label{eq:mozg egyenlet rendenk=0000E9nt}
\end{equation}
with the gauge constraint

\begin{equation}
\partial^{\mu}\tilde{h}_{\mu\nu}^{GW}=0\rightarrow\partial^{\mu}\tilde{h}_{\mu\nu}^{(0)GW}=0,\,\,\,\partial^{\mu}\tilde{h}_{\mu\nu}^{(1)GW}=0.
\end{equation}

We can rewrite Eq. (\ref{eq:mozg egyenlet rendenk=0000E9nt}) order by order as

\begin{eqnarray}
\Box h_{\mu\nu}^{(0)\,GW} & = & -\kappa T_{\mu\nu}\label{eq:perturb mozg egyenlet GW-re}\\
\Box h_{\mu\nu}^{(1)GW}+2\Lambda h_{\mu\nu}^{(0)GW} & = & 0.\nonumber
\end{eqnarray}

The first equation and its gauge condition are the same in Lorenz gauge. The residual symmetry provides a connection between the waves propagating in the Minkowski space and $h^{(0)GW}$: both satisfy the same equations so we can use the familiar results of gravitational waveforms calculated in the usual TT gauge \cite{key-4}. The second equation can be used to study the effect of $\Lambda$ on the gravitational wave.

\subsection{Choose of coordinate system}

To construct coordinates with proper physical meaning, first we have to specify coordinates, in which $\Box\tilde{h}_{\mu\nu}=0$ are fulfilled in the $\Lambda$-gauge. We can use (\ref{eq:h megold=0000E1sa Lambda gaugeban}) to reach the invariant line element

\begin{eqnarray}
ds^{2} & = & \left[1+\frac{\Lambda}{9}\left(3t^{2} - 2r^{2}\right)\right]dt^{2}-\left[1-\frac{\Lambda}{9}\left(-2t^{2}
+ 2 r^{2}+x_{i}^{2}\right)\right]dx_{i}^{2}\nonumber\\ && -  \frac{2\Lambda}{9}tx_{i}dtdx_{i}
+\frac{2\Lambda}{9}x_{i}x_{j}dx_{i}dx_{j}
\end{eqnarray}
and we can transform this expression to the static solution of \cite{key-2}

\begin{equation}
ds^{2}=\left[1-\frac{\Lambda}{3}r'^{2}\right]dt'^{2}-\left[1-\frac{\Lambda}{6}\left(r'^{2}+3x_{i}'^{2}\right)\right]dx_{i}'^{2}
\end{equation}
by using the most general form of (\ref{eq:atvaltas x->x' kozott}), with the following transformation

\begin{eqnarray}
x & = & x'+\frac{\Lambda}{9}\left(-t'^{2}-\frac{x'^{2}}{2}+\frac{\left(y'^{2}+z'^{2}\right)}{4}\right)x'\nonumber \\
y & = & y'+\frac{\Lambda}{9}\left(-t'^{2}-\frac{y'^{2}}{2}+\frac{\left(x'^{2}+z'^{2}\right)}{4}\right)y'\label{eq:elso atvaltas}\\
z & = & z'+\frac{\Lambda}{9}\left(-t'^{2}-\frac{z'^{2}}{2}+\frac{\left(x'^{2}+y'^{2}\right)}{4}\right)z'\nonumber \\
t & = & t'-\frac{\Lambda}{18}\left(t'^{2}+r'^{2}\right)t'.\nonumber
\end{eqnarray}
An additional coordinate transforation

\begin{eqnarray}
x' & = & x"+\frac{\Lambda}{12}x"^{3}\nonumber \\
y' & = & y"+\frac{\Lambda}{12}y"^{3}\label{eq:masodik atvaltas}\\
z' & = & z"+\frac{\Lambda}{12}z"^{3}\nonumber \\
t' & = & t"\nonumber
\end{eqnarray}
allows us to introduce a fully spherically symmetric solution. With

\begin{equation}
r"=\hat{r}+\frac{\Lambda}{12}\hat{r}^{3}\ ,\,\,\,\,\,t"=\hat{t}\label{eq:harmadik atvaltas}
\end{equation}
we can perform another transformation to obtain the Schwarzschild-de Sitter (SdS) metric

\begin{equation}
ds^{2}=\left[1-\frac{\Lambda}{3}\hat{r}^{2}\right]d\hat{t}^{2}-\left[1-\frac{\Lambda}{3}\hat{r}^{2}\right]^{-1}\hat{r}^{2}+\hat{r}^{2}d\Omega^{2}.\label{eq:SdS metrika}
\end{equation}

The linearization of Eq. (\ref{eq:einstein egyenlet, nem perturbalt}) is consistent with this metric, but it is neither homogeneous nor isotropic. This is in contrast with the fundamental principles of cosmology, therefore the Friedmann--Robertson--Walker (FRW) metric should be used instead of the SdS metric \cite{key-1}. The following explicit transformations

\begin{equation}
\hat{r}=\exp\left\{ T\sqrt{\frac{\Lambda}{3}}\right\} R\ ,\,\,\,\,\,\hat{t}=\sqrt{\frac{3}{\Lambda}}\log\left\{ \frac{\sqrt{3}}{\sqrt{3-\Lambda\exp\{2T\sqrt{\frac{\Lambda}{3}}\}R^{2}}}\right\} +T\label{eq:SdS-FRW atvaltas}
\end{equation}
convert (\ref{eq:SdS metrika}) to the FRW metric

\begin{equation}
ds^{2}=dT^{2}-\exp\left\{ 2T\sqrt{\frac{\Lambda}{3}}\right\} d\vec{X}^{2}\label{eq:FRW metrika}
\end{equation}
where the $X^{i}$ are comoving coordinates.

\section{Quadrupole signal}

To study the effects of the $\Lambda$ cosmological constant, especially the $\mathcal{O}(\Lambda)$ part of (\ref{eq:perturb mozg egyenlet GW-re}), we use the quadrupole signal. The plus component of the waveform is written as \cite{key-4}

\begin{equation}
h_{+}=\frac{A(\omega)}{r}\cos\left\{ 2\omega\left(t-r\right)\right\} .
\end{equation}

For simplicity, the amplitude $A(\omega)=\frac{G\mu\omega^{2}R^{2}}{c^{4}}\left(\frac{1+\cos^{2}\iota}{2}\right)$ is set to be $A(\omega)/r=1$ and $\iota=0$.

\subsection{Linearized solution}

To demonstrate the physically relevant effects of the perturbations, we have to apply the coordinate transformations (\ref{eq:elso atvaltas},\ref{eq:masodik atvaltas},\ref{eq:harmadik atvaltas}) and calculate the $h^{(1)GW}$ with Eq. (\ref{eq:perturb mozg egyenlet GW-re}) to find\footnote{In the coordinate transformation the argument of sine and cosine functions are transformed always exactly resulting in an error present in the calculations. At the lowest order in $\Lambda$ this error term is $\leq10^{-3}$, as it can be found in \cite{key-1}. The effects studied here is approximately one order of magnitude higher than this error.}

\begin{equation}
h_{+}^{(1)}=\frac{A(\omega)}{\hat{r}}\left[\frac{\Lambda}{8\omega^{2}}\cos\left\{ 2\omega\left(\hat{t}-\hat{r}\right)\right\} -\frac{\Lambda\hat{r}}{2\omega}\sin\left\{ 2\omega\left(\hat{t}-\hat{r}\right)\right\} \right].
\end{equation}

Finally, to express the results in comoving coordinates, the transformation (\ref{eq:SdS-FRW atvaltas}) has to be applied. Introducing the retarded time $\tau=T-R$, one can derive the following expression for $h_{+}$

\begin{eqnarray}
h_{+} & = & \frac{A(\omega)}{R}\left(1+z_{amp}\right)\cos\left\{ 2\omega\tau\left(1+z_{pha}\right)+\phi\left(R,\Lambda\right)\right\} +\nonumber \\
 &  & \frac{A(\omega)}{R}\left(1+z_{amp}\right)\frac{\Lambda}{8\omega^{2}}\cos\left\{ 2\omega\tau\left(1+z_{pha}\right)+\phi\left(R,\Lambda\right)\right\} \label{eq:h+ minden taggal}\\
 &  & -A(\omega)\frac{\Lambda}{2\omega}\sin\left\{ 2\omega\tau\left(1+z_{pha}\right)+\phi\left(R,\Lambda\right)\right\} +\mathcal{O}(\Lambda^{3/2})\nonumber
\end{eqnarray}
where

\begin{eqnarray}
z_{amp} & = &-\sqrt{\frac{\Lambda}{3}}\left(R-\tau\right)+\frac{\Lambda}{3}\left(R^{2}+\frac{5R\tau}{3}+\frac{5\tau}{6}\right)+\mathcal{O}(\Lambda^{3/2})\\
z_{pha} &= & -\sqrt{\frac{\Lambda}{3}}R-\frac{\Lambda}{3}\left(\frac{2R\tau}{3}-\frac{\tau^{2}}{6}\right)+\mathcal{O}(\Lambda^{3/2})\label{eq:z pha =0000E9s z amp kiirva}
\end{eqnarray}
and $\phi(R,\Lambda)$ is an overall phase factor.

\subsection{Discussion of the results}

In the previous subsection, the most relevant effects of $\Lambda$ have been derived. In order to demonstrate these effects, $h_{+}$ in Eq. (\ref{eq:h+ minden taggal}) is plotted as

\begin{equation}
h_{+}=\frac{A(\omega)}{r}\left(1-\sqrt{\frac{\Lambda}{3}}R\right)\cos\left\{ 2\omega\tau\left(1-\sqrt{\frac{\Lambda}{3}}R-\frac{\Lambda}{3}\frac{2R\tau}{3}\right)\right\} .
\end{equation}

\begin{figure}
\centering
\includegraphics[scale = 0.5]{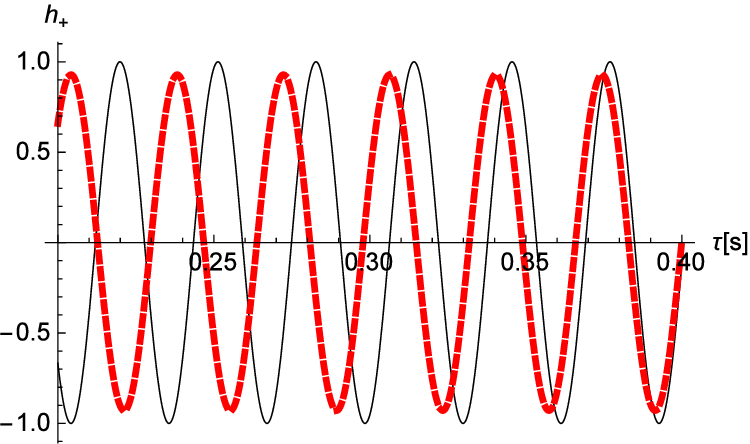}
\caption{\label{fig:1} The $h_+$ waveform in terms of the retarded time $\tau$.The solid line corresponds to $\Lambda=0$ and the dotted line to $\Lambda=10^{-52}\,m^{-2}$.  The distance is $R_{1}=410\,Mpc$. }
\end{figure}

\begin{figure}
\centering
\includegraphics[scale = 0.5]{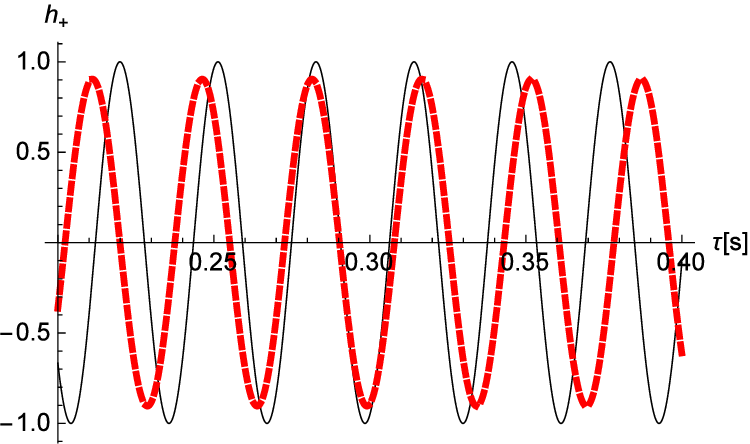}
\caption{\label{fig:2} The $h_+$ waveform in terms of the retarded time $\tau$. The solid line corresponds to $\Lambda=0$ and the dotted line to $\Lambda=10^{-52}\,m^{-2}$. The distance is $R_{2}=570\,Mpc$.}
\end{figure}

\begin{figure}
\centering
\includegraphics[scale = 0.5]{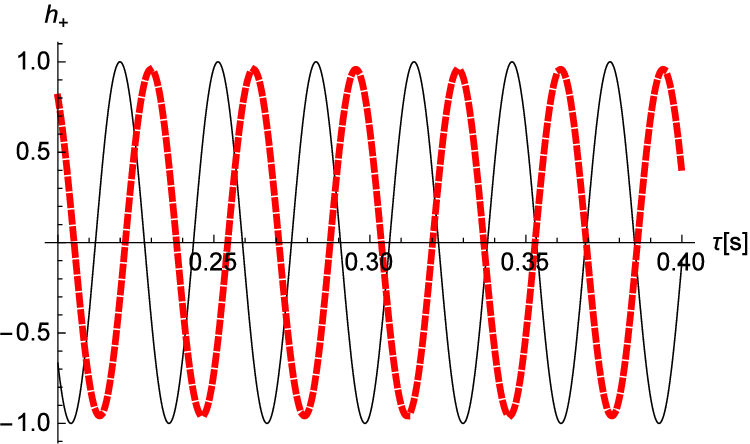}
\caption{\label{fig:3} The $h_+$ waveform in terms of the retarded time $\tau$. The solid line corresponds to $\Lambda=0$ and the dotted line to $\Lambda=10^{-52}\,m^{-2}$. The distance is $R_{3}=230\,Mpc$. }
\end{figure}

In Figs. \ref{fig:1}-\ref{fig:3} $\omega=100\,s^{-1}$ was chosen to analyse the effects of $\Lambda$ during a few cycles. The dependency of the amplitude and wavelength on $\tau$ for constant values of $R$ is shown in Figs. \ref{fig:1}-\ref{fig:3}. These distances are in accordance with those of the event GW150914 where the luminosity distance was $R=410\,(+160\,-180)\,Mpc$~\cite{key-6}.

\begin{figure}
\centering
\includegraphics[scale = 0.5]{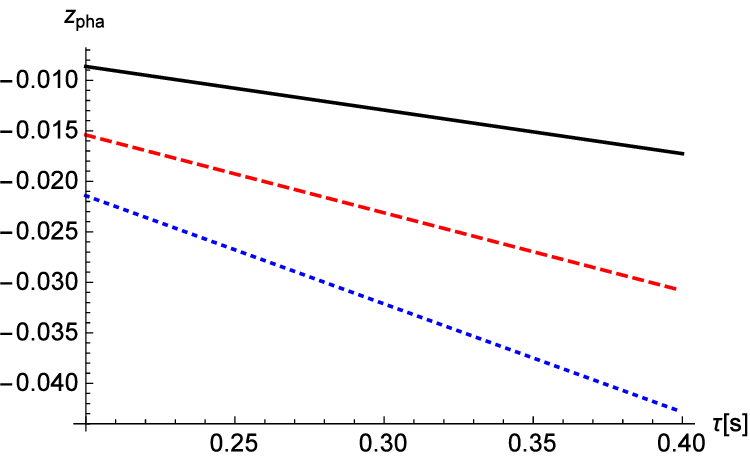}
\caption{\label{fig:4} Dependence of $z_{pha}$ on $\tau$ in the same interval. The solid line corresponds to $R=230\,Mpc$, the dashed line to $R=420\,Mpc$ and the dotten line to $R=570\,Mpc$.}
\end{figure}

The extra cosine and sine functions on the r. h. s. of (\ref{eq:h+ minden taggal}) do not have relevant effects. Setting $R=410\,Mpc$ and $\Lambda=10^{-52}\,m^{-2}$, the cosine and the sine functions will be relevant ($\sim1\%$ of the amplitude of the original magnitude) at $\omega\approx10^{-17}\,s^{-1}$. These effects could be substantial in the theoretical approach where the exact form of $\Lambda$ is left in the waveform. We have also plotted the $z_{pha}$ of (\ref{eq:z pha =0000E9s z amp kiirva}) in the Fig. \ref{fig:4}.

The effects of $\Lambda$ on the $z_{amp}$ is independent of the different values of $\tau$. Indeed, the function $z_{amp}$ is determined by $R$ and $\Lambda$. Consequently, for $R=570,420,230\,Mpc$, $z_{amp}$ has the following values: $-0.095$,$-0.071$ and $-0.042$, respectively.

\section{Summary}

The approach presented in \cite{key-1} can be used to derive observably relevant effects of a non-zero cosmological constant. The perturbative approximation in the presence of $\Lambda$ could be relevant for a quadrupole signal or the detected gravitational waves, so the value of $\Lambda$ is always explicitly kept except in the figures. Due to the coordinate transformations considered here perturbation terms proportional to $\mathcal{O}(\Lambda)$ appear. The transformation to FRW coordinates results in the appearance of additional $\mathcal{O}(\sqrt{\Lambda})$ terms. These modifications could be relevant to specify the property of the detected GW sources. As an extension of the present analysis other types of waveforms, especially the chirp signal \cite{key-4} could be studied. In this paper we have studied a different kind of wave signal compared to the results of \cite{key-1,key-2} where the authors focused on the $E_{\mu\nu}\cos(kx)+D_{\mu\nu}\sin(kx)$ wave-like solution in order to discuss the perturbation method. Though the effect is small, the presence of the cosmological constant modifies both the phase and the amplitude of the quadrupole signal, it is red-shifted and the amplitude of waves decrease as the retarded time $\tau$ grows.

\section*{Acknowledgements}

This work was supported by the NKFIH 124366 and NFKIH 124508 grants. L. \'A. S. has been supported by the GPU Laboratory of Hungarian Academy of Sciences Wigner Research Centre of Physics.  Partial support comes from NewCompStar, COST Action Program MP1304.

%\begin{thebibliography}{000} %for 3 digits

\end{document}